\documentclass[10pt, letter, twocolumn]{IEEEtran}

\usepackage{graphicx}
\usepackage{epsfig}
\usepackage{algorithm}
\usepackage{algorithmic}
\usepackage{ifmtarg}
\usepackage{caption}
\usepackage{footnote}
\usepackage{cite}
\usepackage{amssymb}
\usepackage{amsthm}
\usepackage[fleqn]{amsmath}
\usepackage{amsmath}
\usepackage{amsfonts}
\usepackage{epstopdf}
\usepackage{enumitem}
\usepackage{caption}
\usepackage{subcaption}
\usepackage{multirow}
\usepackage{rotating}
\usepackage{color}
	
\usepackage[utf8]{inputenc}
\usepackage[english]{babel}

\usepackage[font=normalsize]{caption}
\begin{document}
\title{Resource and Placement Optimization for Multiple UAVs using Backhaul Tethered Balloons}
\author{ \IEEEauthorblockN{Ala Alzidaneen, Ahmad Alsharoa, and Mohamed-Slim Alouini \vspace{-0.2cm}}
\thanks{Ala Alzidaneen and Ahmad Alsharoa are with Missoui University of Science and Technology (Missouri S\&T), Rolla, Missouri, United States. E-mail: \{aalzidaneen, aalsharoa\}@mst.edu.

Mohamed-Slim Alouini is with King Abdullah University of Science and Technology (KAUST), Thuwal, Makkah Province, Saudi Arabia. E-mails: slim.alouini@kaust.edu.sa.
}\vspace{-1cm}}

\maketitle
\thispagestyle{empty}
\pagestyle{empty}

\begin{abstract}
\boldmath{
This paper studies the improvement of the achievable end-to-end data rate of ground users assisted with unmanned aerial vehicles (UAVs) and tethered balloons (TBs).
The goal is to maximize the end-to-end throughput of a network suffering from the absence of terrestrial infrastructure.
First, we solve an integer linear programming problem to optimize the associations. Then, we solve the UAVs’ transmit powers optimally by converting the problem into a convex one. Subsequently, an efficient algorithm is proposed to optimize the UAVs' placement. Finally, our mathematical formalism is illustrated with some selected numerical results that show the advantages provided by our proposed scheme.
}
\end{abstract}
\vspace{-.7cm}
\section{Introduction}\label{Introduction}
The use of unmanned aerial vehicles (UAVs) to assist cellular networks has recently attracted a considerable attention due to their dynamic and quick deployment together with their relative low cost~\cite{UAV_mag}. Compared to classical terrestrial networks, using UAVs as flying base stations can be a more effective solution in infrastructure less situations by enabling on-demand throughput~\cite{UAV_mag}. 
Using UAVs as aerial base stations comes with its own challenges.
The UAVs' placement optimization is considered one of the main challenges in UAV communications, especially in the case of multiple UAVs serving multiple users~\cite{pathdeployment3}. In~
In~\cite{zhang1}, the authors investigated the trajectory optimization of one UAV, taking into consideration the energy consumption of the UAV and the users' throughput. 
Furthermore, the multiple UAV placement with energy management has been investigated in~\cite{TMC2019}. The goal was to minimize the overall network energy consumption by proposing an efficient switching on/off technique for terrestrial base stations, depending on the user's requirements.
The work in \cite{Mozaffari1} proposed a cache-enabled UAV framework that enable cloud
radio access networks to meet the users' throughput requirements. The authors objective was to maximize the quality-of-experience of users in the cloud while minimizing the UAVs' transmitting powers.
The work in~\cite{Mozaffari2} proposed a framework that integrates multiple layers of cellular networks such as low and high altitude platforms. In addition to resource managements, the authors in~~\cite{Mozaffari2} introduce an approach based on truncated octahedron cells that determines the minimum number of UAVs needed to cover certain 3D space.

Optimizing the UAVs' placement with resource allocations and associations between UAVs and users can significantly improve the network performance by extending the covering areas and the UAVs' battery operating times. 
The authors in~\cite{zhang2} proposed a one dimensional trajectory solution of one UAV serving multiple users using time division multiple access. 
The work in~\cite{ICCW_2018} considers multiple UAVs with trajectory optimization where the UAVs work as relays that broadcast the signal from certain users to one predefined destination. The goal of that study was to modify the pre-defined path of the UAVs with some threshold in order to achieve higher users' throughput.
In addition to the placement challenge, the channel modeling is another important factor. For instance, the work in \cite{Bor1} proposed a channel model that considers more information about the environment such as the city layout, blocking/obstacle information, reflection, and propagation.
In~\cite{Bor2}, Bor \textit{et al.}  proposed a spatial network scheme that uses UAVs for access point configurations. That work also showed joint spatial network configuration that consists of 3D placement and incentive design for the user-in-the-loop.

Considering the backhauling links of multiple UAVs will enhance the end-to-end throughput but on the other hand, it will also add more challenges. This will  significantly effect the UAV's placement, resource allocation, and associations.
Few works in the literature consider the backhauling connection. For instance, in~\cite{Li2018}, the authors propose multi-hop backhaul connections scheme among multiple UAVs. The authors consider employing multiple UAVs as a quasi-stationary aerial base station to serve multiple users that are distributed on the ground and provide reliable connections to the core network in multi-hop connections.
However, our work propose a more robust solution by considering only one hop from UAVs to tethered balloons (TBs). In addition, we propose to have TBs at a higher altitude than UAVs, which will increas the probability of having line-of-sight (LoS). 
In this paper, we consider a practical scheme for infrastructure-less environment by proposing multiple UAVs that establish backhauling links with TB. 
Given the TBs' locations, we propose to optimize UAVs' placement, resource allocation, and associations. Note that it is not only the channel gains between the UAVs and users need to be enhanced, but also the channel gains between the TBs and UAVs in order to improve the end-to-end throughput.
To the best of the authors knowledge, this is the first work that optimizes the resource allocation and association of multiple UAVs system assisted with TBs for backhauling.
The main contribution in our paper can be summarized as:
\begin{itemize}
\vspace{-.2cm}
\item Formulate an optimization problem aiming to maximize the users' utility in an infrastructure-less environment taking into consideration not only the access link constraints between UAVs and users, but also the backhauling link constraints between UAVs and TBs.
\item Due to non-convexity of the formulated problem, we propose an efficient three-steps solution. Firstly, we solve an integer linear programming (ILP) optimization problem to optimally determine the UAVs' associations (access associations between UAVs and users and backhauling associations between UAVs and TBs).
Then, given the optimal associations, we solve optimally the transmit powers of the UAVs by converting the problem into a convex one.
In the last step, we propose an efficient and low complexity heuristic algorithm based on shrink-and-realign (S\&R) process to optimize the UAVs' placement.
\item The performance of our proposed solution is compared to two benchmarks scenarios: (a) optimizing only the association with fixed UAVs' transmit powers, and (b) random association with fixed UAVs' transmit powers.
\end{itemize}


\vspace{-0.5cm}
\section{System Model}\label{SystemModel}

\begin{figure}[h!]
  \centerline{\includegraphics[width=2in]{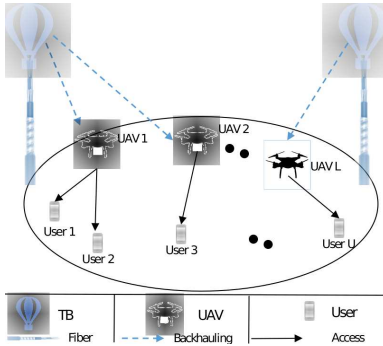}}
   \caption{System model.}\label{system}
   \vspace{-0.3cm}
\end{figure}

We consider a wireless system consisting of $L$ UAVs aiming to provide downlink data to $U$ users in a certain geographical area as shown in~Fig.~\ref{system}.
We also consider $M$ TBs placed in fixed locations and connected directly to fiber links to be used for the UAV backhauling, where each UAV can be associated with one TB.
We consider a central unit located at one of the TBs to manage the resources. The central unit can exchange the decision messages with the other TBs (i.e., it can via radio frequency band or higher bands such as optical band). Then, each TB exchanges the control messages with the associated UAVs.
Let us consider a 3D coordinate system where the coordinate of TB $m$, UAV $l$, and user $u$ are given, respectively, as $J_m^\mathcal{M}=[x_m^{\mathcal{M}}, y_m^{\mathcal{M}}, z_m^{\mathcal{M}}]^t$ and $J_l^\mathcal{L}=[x_l^{\mathcal{L}}, y_l^{\mathcal{L}}, z_l^{\mathcal{L}}]^t$, $J_u^\mathcal{U}=[x_u^{\mathcal{U}}, y_u^{\mathcal{U}}, 0]^t$, where $[.]^t$ denotes the transpose operator.

To avoid the loop interference, we assume that no interference between the backhauling and access links. Therefore, the backhauling and access links should operate on orthogonal resources, meaning that, the available bandwidth is divided sparsely between these two transmission links~\cite{OFDM2}.
In addition, we assume that the UAVs adopt an orthogonal frequency division multiple access (OFDMA) technique to provide service to ground users. We assume that the available spectrum is divided into $N$ resource blocks (RBs). Each RB has a bandwidth of $B= 180$ KHz~\cite{3GPP_release11a}. Therefore, there will be no intra-cell and inter-cell interference on the downlink between the users as they are using orthogonal RBs.
\vspace{-0.3cm}
\subsection{Channel Model}
We distinguish between two channel models depending on the access or backhauling communication link.

\subsubsection{Acess Link: UAV-to-User Channel Gain}
As discussed in~\cite{PL1}, the user can receive two main signals from UAVs. The first one is the LoS signal and the second one is a strong reflected non line-of-sight (NLoS) signal. These types can be considered separately with different probability of occurrence (PoO).
By considering the LoS and NLoS component with their PoO separately, the access channel gain between UAV $l$ and user $u$ over RB $n$ is given as~\cite{PL1}
\begin{equation}\label{front}
h^A_{lu,n}=(PL_{lu,n})^{-1}
\end{equation}
In~\eqref{front}, $PL_{lu,n}$ is the average PL contains the probability of the LoS and NLoS links between UAV $l$ and user $u$ over RB $n$ and given by $PL_{lu,n}=p^{\text{LoS}}_{lu} PL^{\text{LoS}}_{lu}  +(1-p^{\text{LoS}}_{lu}) PL^{\text{NLoS}}_{lu},$
where $PL^{\text{LoS}}_{lu,n}$, and $PL^{\text{NLoS}}_{lu,n}$ are the PL between UAV $l$ and user $u$ over RB $n$ in for LoS and NLoS, and given, respectively, as
\small
\begin{equation}\label{LoS}
PL^{\text{LoS}}_{lu,n}= \xi_{\text{LoS}} \left(\frac{4 \pi \alpha_{lu} f_c}{C}\right), PL^{\text{NLoS}}_{lu,n}= \xi_{\text{NLoS}}\left(\frac{4 \pi \alpha_{lu} f_c}{C}\right),
\end{equation}
\normalsize
where $\alpha_{lu}$ is the distance between UAV $l$ and user $u$. $C$ and $f_c$ are the speed of light and the radio signal carrier frequency, respectively. The parameters $\xi_{\text{LoS}}$ and $\xi_{\text{NLoS}}$ are the additional loss to the free space propagation for LoS and NLoS links, respectively, due to the shadowing effect and the reflection of signals from obstacles.
$p^{\text{LoS}}_{lu}$ is the LoS probability and given as $p^{\text{LoS}}_{lu}=1/(1+c_1 \exp(-c_2[\theta_{lu}-c_1])) $~\cite{PL1},
where $\theta_{lu}=(180/\pi) \sin^{-1}\left(z_l^{\mathcal{L}}/\alpha_{lu}\right)$ is the elevation angle between the UAV $l$ and user $u$ in degree. The constants $c_1$ and $c_2$ are values depend on the environment. Thus, NLoS probability is equal to $1-p^{\text{LoS}}_{lu}$.
\subsubsection{Backhauling: TB-to-UAV Channel Gain}
The channel gains between TBs and UAVs are depending on large-scale and small-scale fading. The large-scale fading is a result of free space path loss and attenuation due to some environmental effects
On the other hand, the small-scale fading can be modeled as Ricean fading due to the presence of LoS links from the TBs to UAVs. Therefore, the channel gain between TB $m$ and UAV $l$ over RB $n$ can be given as shown in~\cite{HAP_channel2}:
\begin{equation}\label{channel_back}
h^B_{ml,n}=\left(\frac{C}{4\pi \beta_{ml} f_c}\right)^2 \Phi_{ml,n},
\end{equation}
where $\beta_{ml}$ is the distance between TB $m$ and UAV $l$. The parameter $\Phi_{ml,n}$ is the Rician small-scale gain between TB $m$ and UAV $l$ over RB $n$ with Rician factor equal to $\kappa$.

\vspace{-0.3cm}
\subsection{Associations}
We distinguish between two type of associations: the access association between UAVs and users and backhauling association between TBs and UAVs.
\subsubsection{Access Link Association}
We introduce a binary variable $\epsilon_{lu,n}$ that indicates the association between UAV $l$ and user $u$ over RB $n$, where $\epsilon_{lu,n}=1$ if UAV $l$ is associated with user $u$ over RB $n$ and $0$, otherwise. 
We assume that multiple users can be associated with one UAV over different RBs. On the other hand, each use is allowed to associate with only one UAV. On the other hand, each user should be associated with a unique RB at most
Therefore, the following constraints need to be respected:
\begin{equation}
\sum\limits_{l=1}^L \sum\limits_{n=1}^N \epsilon_{lu,n} \leq 1,  \quad \forall u, \quad \sum\limits_{l=1}^L \sum\limits_{u=1}^U \epsilon_{lu,n} \leq 1,  \quad \forall n.
\end{equation}


\subsubsection{Backhauling Link Association}
For the backhauling, each UAV should be strictly associated with only one TB. Therefore, we introduce another binary variable, $\vartheta_{ml}$,
where it is equal to 1 if TB $m$ is associated with UAV $l$ and 0 otherwise. Hence, the following equality needs to be respected:
\begin{equation}
 \sum\limits_{m=1}^M \vartheta_{ml} = 1, \quad \forall l.
\end{equation}

\section{Problem Formulation}\label{ProblemFormulation}
The downlink access data rate between UAV $l$ and user $u$ over RB $n$ can be expressed as
\begin{equation}\label{Rate_front}
R_{lu,n}=B \log_2\left(1+\frac{P_{lu,n} h^F_{lu,n}}{B N_0}\right),
\end{equation}
where $B$ and $N_0$ are the transmission access bandwidth and the noise power, respectively. Note that we assume all the access transmissions between the UAVs and users operate sparsely (i.e., allocating different RBs to different users), hence, there is neither intra-cell nor inter-cell interference between users). We plan to study the cross-interference cases in a future extension of this work.
For simplicity and without loss of generally, we assume the backhauling bandwidth and the
transmit power of TB $m$ are uniformly distributed over all UAVs.
Therefore, the backhauling transmission rate from TB $m$ to UAV $l$ can be expressed as
\begin{equation}\label{Rate_back}
R_{ml}= B_0 \log_2\left(1+\frac{P_0 h^B_{ml}}{B_0 N_0}\right).
\end{equation}
where, $B_0$, $P_0$ are the uniform bandwidth and transmit power associated with the backhauling link from TB $m$ to UAV $l$.

In the sequel, we formulate an optimization problem aiming to maximize the end-to-end throughput by optimizing the following:
1) backhauling link association between TBs and UAVs ($\vartheta_{ml}$), 2) access link association between UAVs and users ($\epsilon_{lu,n}$), 3) access transmit powers of the UAVs ($P_{lu,n}$), and 4) UAVs' placement ($J_l^\mathcal{L}$). Therefore, our optimization problem can be now formulated as:
\begin{align}
&\hspace{-0.7cm}\textbf{(\text{P0}):} \quad \underset{\substack{\vartheta_{ml}, \epsilon_{lu,n}, \\ P_{lu,n}, J_l^\mathcal{L} }}{\text{maximize}}
\sum\limits_{l=1}^L \min \left( \sum\limits_{u=1}^U \sum\limits_{n=1}^N  \epsilon_{lu,n} R_{lu,n}, \sum\limits_{m=1}^M \vartheta_{ml} R_{ml} \right) \label{of}\\
&\hspace{-0.5cm}\text{subject to:}\nonumber\\
&\hspace{-0.5cm}\sum\limits_{u=1}^U \sum\limits_{n=1}^N \epsilon_{lu,n} P_{lu,n} \leq \bar{P}_l,  \quad \forall l,\label{power_r}\\
&\hspace{-0.5cm} \sum\limits_{l=1}^L \sum\limits_{n=1}^N \epsilon_{lu,n} \leq 1,  \quad \forall u, \label{epsilon1}\\
&\hspace{-0.5cm} \sum\limits_{l=1}^L \sum\limits_{u=1}^U \epsilon_{lu,n} \leq 1,  \quad \forall n,\label{epsilon2}\\
&\hspace{-0.5cm} \sum\limits_{m=1}^M \vartheta_{ml} = 1, \quad \forall l, \label{epsilon3}
\end{align}
where constraint~\eqref{power_r} ensures that the total transmit power of each UAV is limited by the peak power $\bar{P}_l$.
Constraints~\eqref{epsilon1}-\eqref{epsilon3} indicate the access and backhauling association constraints as explained in Section~\ref{SystemModel}-C.

\section{Proposed Solution}\label{ProblemSolution}
The problem $\textbf{\text{P0}}$ is a mixed integer non-linear programming (MINLP) and solving it is a challenging task.
In this section, we propose to solve our optimization problem using a three-steps iterative approach. Firstly, given the initial locations of the UAVs, we propose to find the optimal access and backhauling associations given uninform power distributions. Then, we derive the optimal transmit powers (i.e., $P_{lu,n}$) for these given associations.
Finally, in the third step, given these associations and UAVs' transmit powers, we propose an efficient algorithm to optimize the placement of the UAVs in order to achieve a better end-to-end throughput.

\vspace{-.3cm}
\subsection{Access and Backhauling Association and Power Optimization with Given UAVs Locations}\label{step1}
The goal is to solve the optimization problem with given UAVs locations $J_l^\mathcal{L}$, $\forall l=1,..,L$.
Therefore, $\textbf{\text{P0}}$ can be now simplified as:
\begin{align}
&\hspace{-0.7cm}\textbf{(\text{P1}):} \quad \underset{\substack{\vartheta_{ml}, \epsilon_{lu,n}, \\ P_{lu,n}, R_l }}{\text{maximize}}
\sum\limits_{l=1}^L R_l \label{of1}\\
&\hspace{-0.5cm}\text{subject to:}\nonumber\\
&\hspace{-0.5cm} \sum\limits_{u=1}^U \sum\limits_{n=1}^N  \epsilon_{lu,n} R_{lu,n} \geq R_l, \forall l,  \label{min11} \\
&\hspace{-0.5cm} \sum\limits_{m=1}^M \vartheta_{ml} R_{ml}  \geq R_l, \forall l, \label{min12} \\
&\hspace{-0.5cm} \text{Constraints:}\quad \eqref{power_r}, \eqref{epsilon1}, \eqref{epsilon2}, \eqref{epsilon3}.
\end{align}

Let us start by fixing the transmit power and optimizing the associations. It can be seen that the problem becomes linear in $\vartheta_{ml}$ and $\epsilon_{lu,n}$, and thus it can be solved using on-the-shelf softwares such as the Gurobi/CVX interface.

Next, for the given access and backhauling associations ($\epsilon_{lu,n}$, $\vartheta_{ml}$), problem $\textbf{\text{P1}}$ becomes convex in terms of $P_{lu,n}$ and $R_L$ since the objective function and constraints are convex with respect to both $P_{lu,n}$ and $R_l$.
Therefore, we can solve our convex optimization problem by exploiting its strong duality to find the Lagrangian multipliers that minimize the dual problem as follows:
\begin{equation}
\underset{\boldsymbol{\lambda},\boldsymbol{\mu} \geq 0}{\text{min}}\quad \underset{P_{lu,n}, R_l \geq 0}{\text{max}}\quad \mathcal{L}(\boldsymbol{\lambda},\boldsymbol{\mu}, P_{lu,n}, R_l),
 \end{equation}
 The Lagrangian function $\mathcal{L}$ can be derived as:
\small
\begin{align}\label{lagr}
&\hspace{-0.5cm} \mathcal{L}= \sum\limits_{l=1}^L R_l - \sum\limits_{l=1}^L \lambda_l \left( \sum\limits_{u=1}^U \sum\limits_{n=1}^N \epsilon_{lu,n} P_{lu,n} - \bar{P}_l \right) + \nonumber \\
&\hspace{-0.5cm} \sum\limits_{l=1}^L  \mu_l \left( \sum\limits_{u=1}^U \sum\limits_{n=1}^N  \epsilon_{lu,n} R_{lu,n} - R_l \right),
\end{align}
\normalsize
where $\boldsymbol{\lambda}=[\lambda_1,..,\lambda_L]$ and $\boldsymbol{\mu}=\mu_1,..,\mu_L$, are the Lagrangian vectors including the Lagrangian multipliers related to constraints~\eqref{power_r} and~\eqref{min11}, respectively. Now, the optimal transmit power can be found by taking the first derivative of~\eqref{lagr} with respect to the $P_{lu,n}$. Hence, the optimal $P^*_{lu,n}$ is given by:
\begin{equation}\label{power_optimal}
P_{lu,n}=\left[\frac{\mu_l B}{\ln(2) \lambda_l} - \frac{B N_0}{h^F_{lu,n}} \right]^+,
 \end{equation}
where the operation $[x]^+$ is the maximum value between $x$ and 0.
After obtaining $P_{lu,n}$ corresponding to the initial values of Lagrangian multipliers, we can employ the subgradient method to find the optimal values~\cite{subgradient}. Hence, to obtain the solution, we start with any initial values for the Lagrangian multipliers and evaluate the optimal $P_{lu,n}$. Then, we can update the Lagrangian multipliers at the next iteration $(i+1)$ according to the non-summable diminishing step size policy given in~\cite{subgradient}. The updated values of the optimal solution and the Lagrangian multipliers are repeated until convergence to find the best optimal solution for~$P^*_{lu,n}$.

\subsection{UAV Placements with Given UAVs' Transmit Powers and Associations}\label{step2}

Due to the non-convexity of optimization problem $\textbf{\text{P0}}$ even with fixed UAVs' transmit power values and associations, we introduce a low complexity and efficient algorithm based on the S\&R process. The proposed algorithm has many advantages over other heuristic algorithms proposed in the literature such simple implementation which results in low complexity and quick convergence to a near optimal solution. Note that our purposed S\&R algorithm is a modified version of the recursive random
search (RRS) algorithm described in~\cite{Murat}, where it has been tested on a suite of well-known and difficult benchmark
functions. The results showed that in terms of quickly locating a “good” solution, RRS outperforms other search algorithms, such as multi-start pattern search and controlled random search.

\begin{figure}[h!]
  \centerline{\includegraphics[width=1.5in]{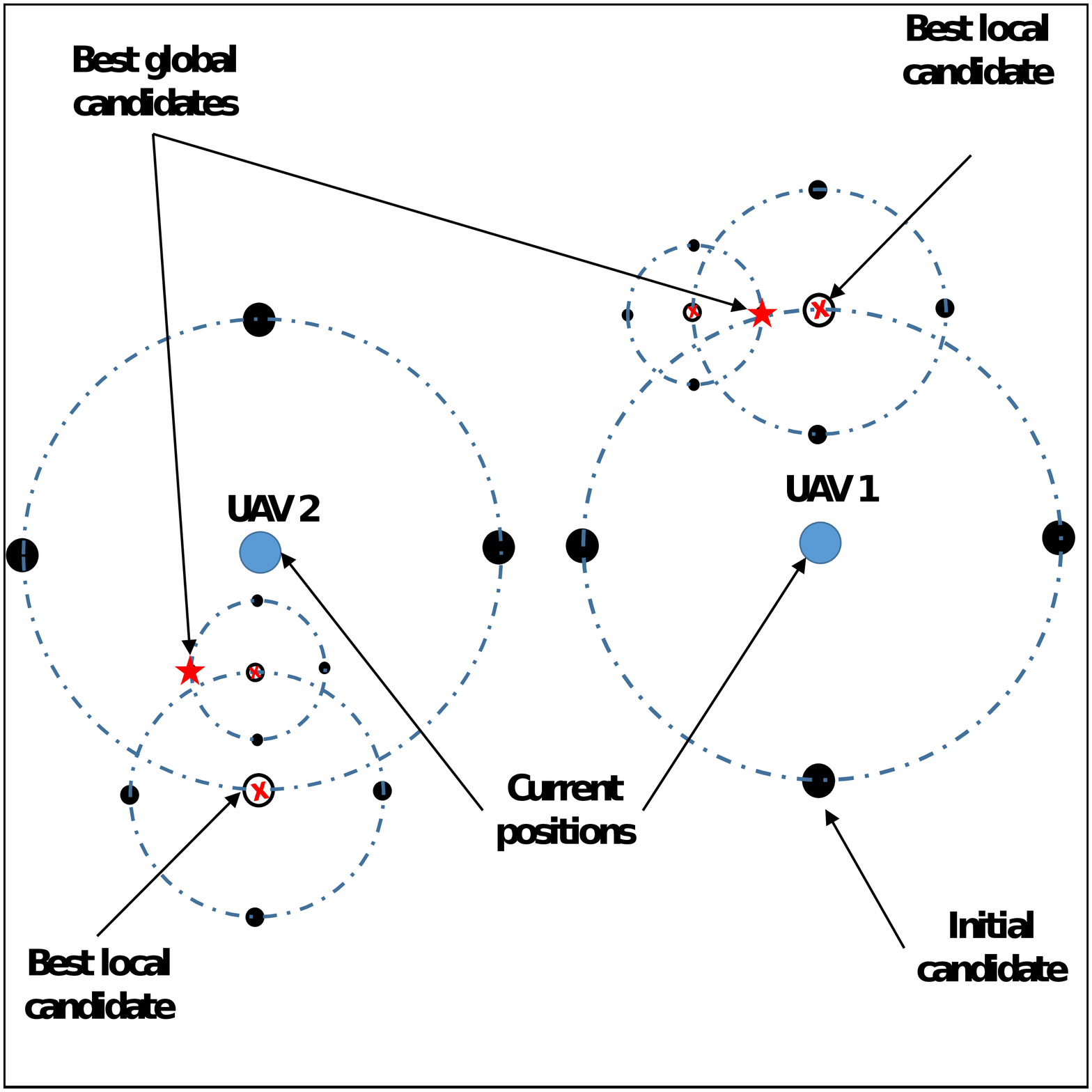}}
   \caption{Our heuristic approach.}\label{SRfig}
\end{figure}

We start our algorithm by generating initial next position candidates $Q_l, \forall l$ as a circle with radius $r(i)$ around each UAV location to form the inial population.
Next, we solve $\textbf{\text{P0}}$ to determine the objective function for each candidates combination. We then find the initial best local candidate combinations $Q^{i,\text{local}}=q_l^{i,\text{local}}, \forall l$ that gives the maximum objective function for iteration $i$.
After that, we apply the S\&R process recursively to find the best global solution $Q^*=q_l^*, \forall l$ by generating a new candidates on a
circle of radius ($\textit{r(i+1)=r(i)/2}$) around each local solution. We repeat this process until the size of the sample space decreases
below a certain threshold or no improvement can be made. Fig.~\ref{SRfig} shows an example of the proposed algorithm using two UAVs ($L=2$) and three maximum iteration.

The details of the joint optimization algorithm that optimizes the placement, transmit power, and associations of the UAVs are given in Algorithm~\ref{joint}.

\begin{algorithm}[h!]
\caption{Joint Placement and Resource Allocation Algorithm}\label{joint}
\small
\begin{algorithmic}[1]
\STATE i=1.
\STATE Generate initial candidates $Q_l, \forall l$ in a circle of radius $r=r(i)$ around each UAV
$J_l^{\mathcal{L}}(q_l),\;l=1 \cdots L$.
\WHILE {{Not} converged or reaching $I_{\max}$}
\FOR {$l=1 \cdots L$}
\FOR {$q_l=1 \cdots Q$}
\STATE Find $\epsilon_{lu,n}$, $\vartheta_{ml}$, and $P_{lu,n}$ by solving $\textbf{\text{P1}}$ as explained in Section. IV-A.
\STATE Compute~\eqref{of}.
\ENDFOR
\ENDFOR
\STATE Find $(q_l^{i,\text{local}})$ that gives the higher objective function combination.
\STATE $r=r(i)/2$
\STATE Start applying S\&R process for the local solution.
\STATE i=i+1.
\ENDWHILE
\end{algorithmic}
\normalsize
\end{algorithm}

\vspace{-0.5cm}
\section{Simulation Results}
In this section, we provide some selected simulation results to demonstrate the benefits of our system model. We consider a system with $U$ users distributed randomly within an area of 1000m $\times$ 1000m. In addition, we consider
$M=2$ TBs at fix location (0,500,200) m and (1000,500,200) m and $L=4$ UAVs flying at a fixed altitude $z_l^{\mathcal{L}}=100$ m $\forall l=1,..,L$. We use $N=30$ available RBs in the fronthuling in our simulations. The maximum transmit peak power of the UAVs is $\bar{P}_l=30$ dBm. The noise power $N_0$ is assumed to be $-110$ dBm.
Table~\ref{tab2} summarizes the remaining parameters that used in the simulations~\cite{TMC2019}.
{\small
\begin{table}[t]
\centering
\caption{\label{tab2} Simulation parameters}
\addtolength{\tabcolsep}{-2pt}\begin{tabular}{|l|c||l|c||l|c|}
\hline
\textbf{Constant} & \textbf{Value} & \textbf{Constant} & \textbf{Value}& \textbf{Constant} & \textbf{Value}\\ \hline \hline
$\lambda$ (m) & 0.125 & $c_1$ & $9.6$ & $c_2$ & $0.29$   \\ \hline
$B$ (kHz) & $180$ & $\xi_{\text{LoS}}$ (dB) & 1 & $\xi_{\text{NLoS}}$ (dB) & 12 \\ \hline
$B_0$ (MHz) & $1$ & $P_0$ (W) & 10 & $\kappa$ & 20 \\ \hline
\end{tabular}
\end{table}
}

\begin{figure}
     \centering
     \begin{subfigure}[b]{0.23\textwidth}
         \centering
         \includegraphics[width=\textwidth]{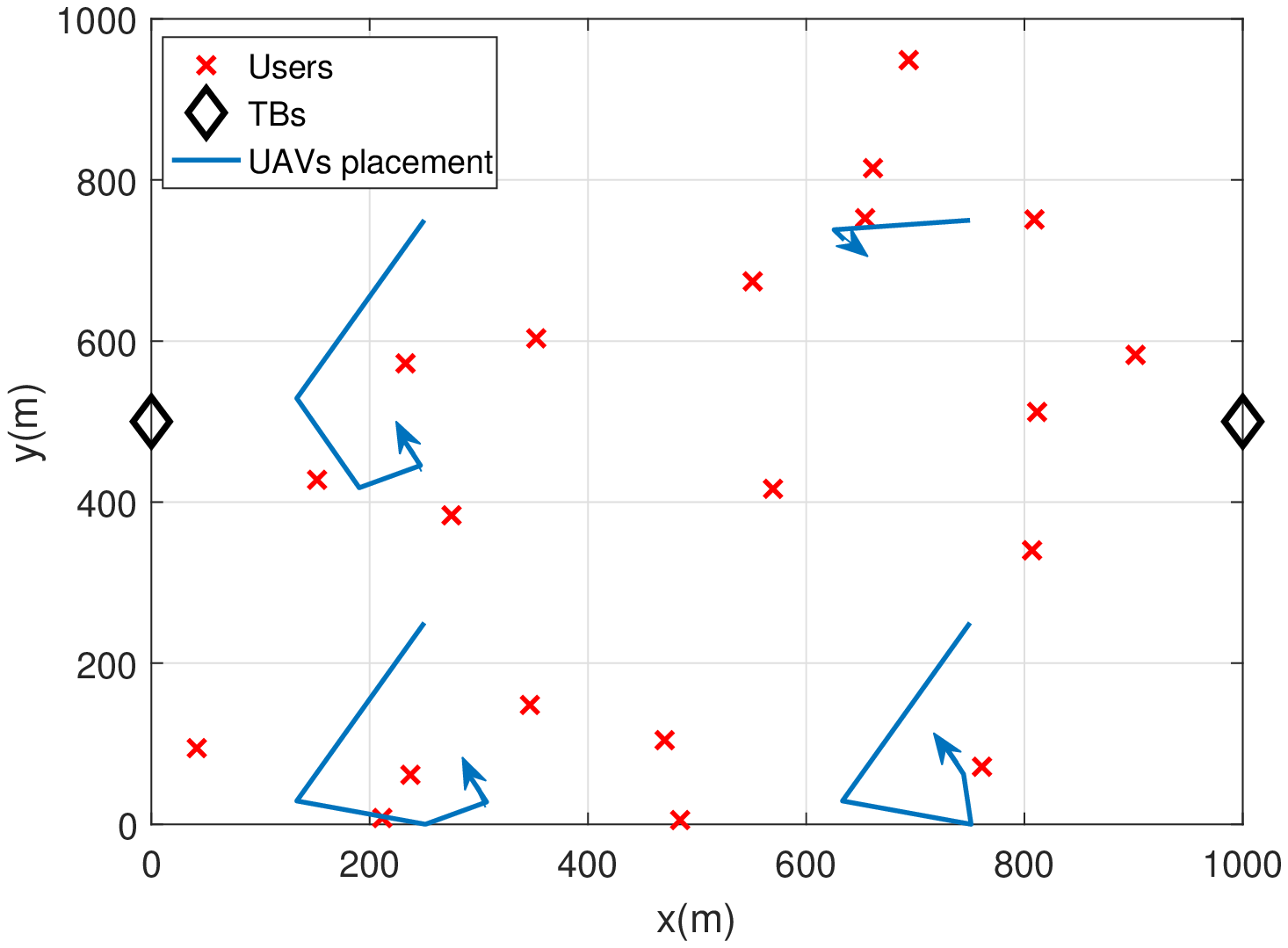}
         \caption{UAVs' placement}
         \label{fig:y equals x}
     \end{subfigure}
     \begin{subfigure}[b]{0.23\textwidth}
         \centering
         \includegraphics[width=\textwidth]{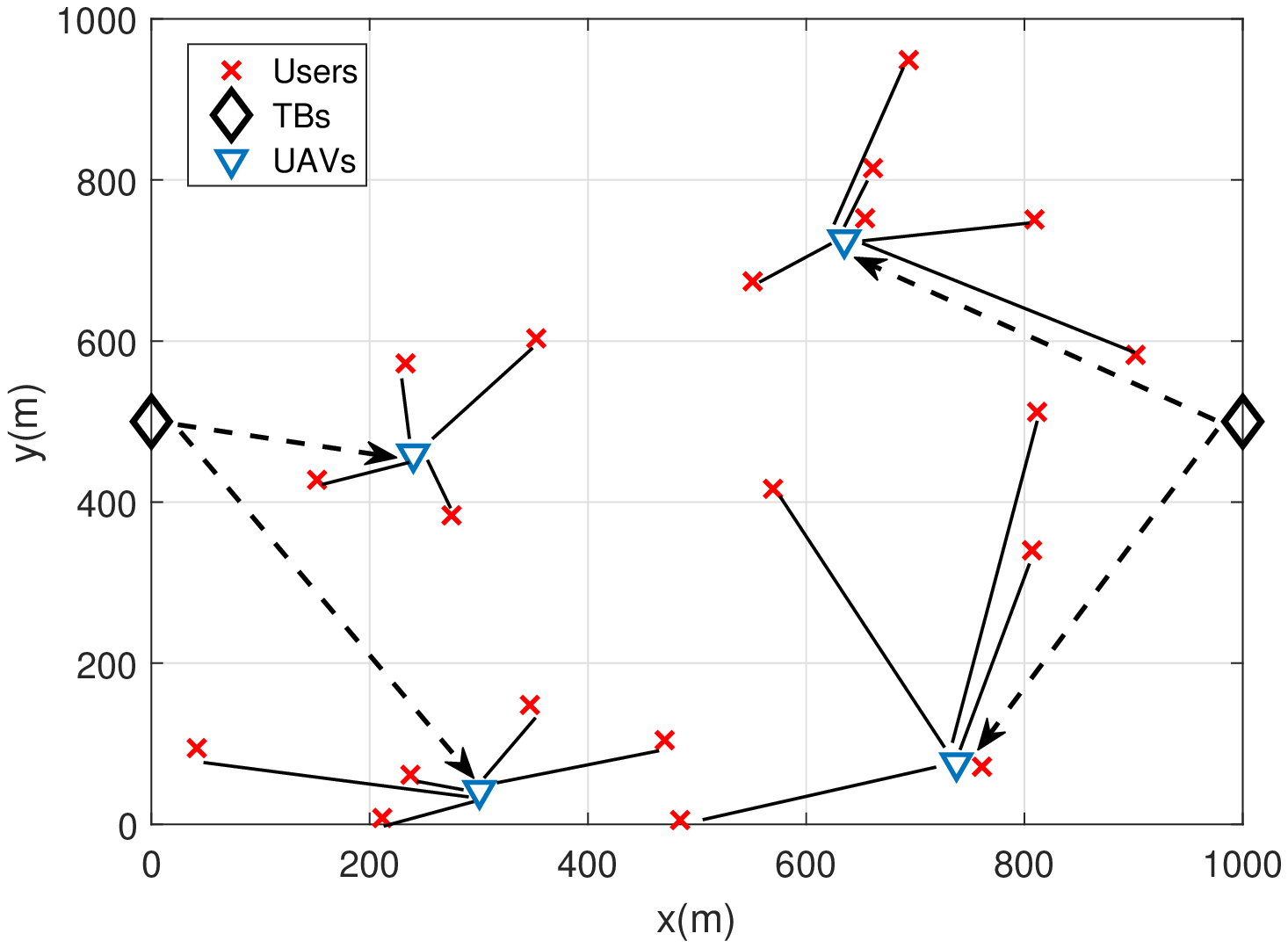}
         \caption{UAVs' association}
         \label{fig:three sin x}
     \end{subfigure}
     \caption{UAVs' placement and association}
        \label{UAV_Placement_association}
\end{figure}

Fig.~\ref{UAV_Placement_association} shows the UAVs' placement and associations for $U=20$ with $\bar{P}_l=30$ dBm. For instance, it can be noticed from Fig.~\ref{UAV_Placement_association} (a) that UAVs need few iterations to reach the near optimal solution. Also, some of the UAVs requires less iteration than others to converge. This can be done when the next candidate position is not better, in terms of data rate, that the current iteration. While Fig.~\ref{UAV_Placement_association} (b) shows the access and backhauling associations. In this case, the UAVs try to find the best placement that satisfied both the access and backhauling links simultaneously.

\begin{figure}[h!]
  \centerline{\includegraphics[width=2in]{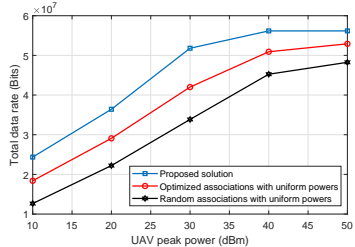}}
   \caption{Total end-to-end users data rate versus UAVs' peak transmit power for $U=20$ users and $B_0=1$ MHz.}\label{Peakpower}
\end{figure}

Fig.~\ref{Peakpower} plots the total achieved end-to-end users' data rate with $U=20$ and $B_0=1$ MHz versus UAVs' peak transmit power. Our proposed solution is compared with two benchmark solutions:
1- Optimizing only the association and the UAVs' placement with uniform power distribution (i.e., $P_{lu,n}=\bar{P}_l/N$),
and 2- Optimizing only the placement of the UAVs with random association and uniform power distribution.
Furthermore, the figure shows that as $\bar{P}_l$ increases, the achievable throughput increases up to a certain value. This can be explained by starting from this point of $\bar{P}_l$, the end-to-end throughput cannot be improved because it depends also on
the backhauling data rate as given in~\eqref{of}. The backhauling data rate works as the bottleneck for the access data rate.
This figure also shows that our proposed algorithm outperforms the other two solutions. For instance, using $\bar{P}_l=30$ dBm, our proposed solution can enhance the end-to-end throughput by around 21\% and 58\% compared to optimizing the association with uniform power and
to random association with uniform power, respectively.
Furthermore, it can be observed that the gap between solutions reduces as $\bar{P}_l$ increases. This is because as $\bar{P}_l$ increases to large values, the effect of uniform power is reduces. However, in practice, the value of $\bar{P}_l$ is limited.

\begin{figure}[h!]
  \centerline{\includegraphics[width=2in]{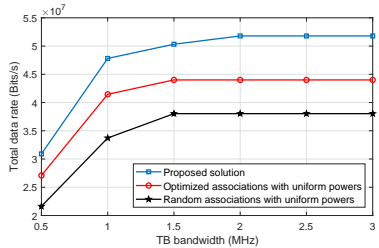}}
   \caption{Total end-to-end throughput versus UAVs' peak transmit power for $U=20$ users and $\bar{P}_l=30$ dBm.}\label{TB_bandwidth}
\end{figure}

On the other hand, Fig.~\ref{TB_bandwidth} shows the total achieved end-to-end users' data rate with $U=20$ and $\bar{P}_l=30$ dBm
versus the UAVs' backhauling bandwidth $B_0$ for different solutions similar to Fig.~\ref{UAV_Placement_association}(a).
This figure shows that the achievable end-to-end throughput is improving with the increase of $B_0$ up to a certain value,
because starting from this point of $B_0$, the achieved end-to-end throughput can not be enhanced further because
it depends on the value of $P_{lu,n}$, which is limited by $\bar{P}_l$ as given in~\eqref{power_r}.

\begin{figure}[h!]
  \centerline{\includegraphics[width=1.9in]{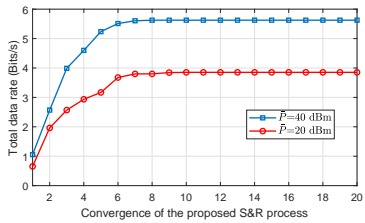}}
   \caption{Convergence of the proposed S\&R process.}\label{convergence}
\end{figure}
Finally, Fig.~\ref{convergence} plots the convergence speed of the S\&R algorithm, which is defined by the number of
iterations needed to reach convergence. Note that one iteration in Fig.~\ref{convergence} corresponds to one iteration of the “while loop” given in Algorithm 1 line 3-14. It can be noted that the algorithm converges within around 6-9 iterations.

\section{Conclusion}\label{Conclusions}
This paper proposed an efficient optimization framework using UAVs as base stations to provide connectivity to the ground users while taking
the backhauling constraints into consideration.
The objective was to maximize the end-to-end throughput by optimizing the placement, transmit power, and associations of the UAVs.
The simulation results illustrated the behavior of our approach and its significant impacts on the end-to-end throughput.
In the next study, a free-space optical (FSO) communication link between UAVs and TB will be considered.
The FSO link in the backhauling not only used to mitigate the bottleneck limitation of the radio frequency link, but also as a harvested source of energy for TBs.
However, it will add more complexity to the problem by optimizing extra parameters such as the LoS angles alignment.

\bibliographystyle{IEEEtran}
\bibliography{J_2019WCL_R1}
\end{document}